\documentclass[aps,pra,twocolumn,amsmath,amssymb,aps,superscriptaddress]{revtex4-1}
\usepackage{amsmath}    
\usepackage{graphicx}   
\usepackage{verbatim}   
\usepackage[colorlinks,urlcolor=blue,citecolor=blue,linkcolor=blue]{hyperref}
\usepackage{color}      
\usepackage{subfigure}  
\usepackage{hyperref}   
\usepackage{dcolumn}
\usepackage{bm}
\usepackage{epsfig}
\usepackage{epstopdf}
\usepackage{braket}
\usepackage{txfonts}

\begin{document}

\title{Tailoring the thermalization time of a cavity-field using distinct atomic reservoirs}

\author{Deniz T\"{u}rkpen\c{c}e}
\affiliation{Department of Electrical Engineering, \.{I}stanbul Technical University, 34469 \.{I}stanbul, Turkey}

\author{Ricardo Rom\'an-Ancheyta}
\email{ancheyta6@gmail.com}
\affiliation{Department of Physics, Ko\c{c} University, 34450 Sar{\i}yer, \.{I}stanbul, Turkey}

\date{\today}
 
\begin{abstract}
We study how the thermalization time of a single radiation cavity-field mode changes drastically depending on the type of the atomic reservoir it interacts. Temporal evolution of the field is analyzed within the micromaser scheme, where each atomic reservoir is modeled  as a beam of atoms crossing an electromagnetic cavity in which they weakly interact with the field. The cavity-field thermalizes when we consider either multi-atom or multi-level atom reservoirs. We found that each atomic reservoir generates a different scaling law in the thermalization time of the cavity-field. Such scaling laws can be used for a faster or slower heating and cooling process. We have obtained analytical expressions for the thermalization time that were verified by means of a numerical simulation of the injection of each atomic reservoir into the cavity. We also discussed how our results could boost the efficiency and power output of some quantum heat engines during a finite time operation when the radiation field mode acts as the working substance.
\end{abstract}


\maketitle

\newcommand{\dbar}{\ensuremath{\mathchar'26\mkern-12mu d}}


\section{Introduction}\label{sec:intro}

A deep understanding of what is heat and work in the quantum domain is one of the main interests and cha\-llen\-ges for the quantum thermodynamics community~\cite{KoslDyna,Janet2016,niedenzu2018quantum} along with the study of heat engines using quantum matter as the working substances~\cite{quan_quantum_2007,kieu_second_2004,quan_quantum_2005,johal_quantum_2009,allahverdyan_work_2008,
wang_efficiency_2012,dorfman_photosynthetic_2013,altintas_quantum_2014,
altintas_rabi_2015,turkpence_quantum_2016,turkpence_photonic_2017}. 
Recently, the first experimental realization of a single atom heat engine has been demonstrated~\cite{rosnagel_single-atom_2016}, there the laser cooling and electric field noise play the role of a cold and a hot reservoirs respectively. In this context, ideas on hybridizing the engineering and quantum physics seems  begin to emerge~\cite{schleier-smith_editorial:_2016}. 

Engineering the states of photonic microwave fields appears to be a requirement for quantum communication and quantum information technologies. Photons are convenient carriers of information with the ability for encoding qubit information into their polarization states, moreover these can be transmitted via optical communication channels~\cite{northup_quantum_2014}. Microwave fields also have an active part in quantum information processing as they govern the mutual interaction between superconducting qubits~\cite{majer_coupling_2007,sillanpaa_coherent_2007}, promising building blocks for quantum computers. However, microwaves are not very efficient for the transmission of information over long distances. A microwave signal has the possibility of being converter into the optical domain using an optomechanical interaction with a micro mechanical resonator~\cite{barzanjeh_reversible_2012,andrews_bidirectional_2014}. To this end, cooling the mechanical resonator to the ground state is an essential step in order to protect it against classical noise~\cite{chan_laser_2011,teufel_sideband_2011}, maybe, a quantum refrigerator~\cite{niedenzu2018quantized} could be used to achieve this task.

Thermalization of quantum systems is a fundamental concept in quantum statistical physics~\cite{kubo_statistical_2003} and has attracted renewed interest. Conventional quantum thermalization states that an open quantum system connected with a thermal reservoir  will evolve towards an equilibrium state with the same temperature of the reservoir in an irreversible way~\cite{breuer_theory_2007}. If the system  is part of an isolated quantum system, then the process can be defined by  the eigenstate thermalization  hypothesis~\cite{deutsch_quantum_1991,srednicki_chaos_1994}. 
Here, we adopt the former conventional view by using atomic reservoirs within the framework of the micromaser model~\cite{filipowicz_theory_1986}.

Beyond quantum optics, micromaser models have become popular in studies of thermalization processes of quantum systems using single atom reservoirs~\cite{liao_single-particle_2010}, coherent or correlated two-atom reservoirs~\cite{li_quantum_2014,dillenschneider_energetics_2009} as well as multi-atom reservoirs~\cite{hardal,dag_multiatom_2016}. Moreover, a micromaser setup was introduced in the seminal paper~\cite{scully_extracting_2003} as model of a quantum heat engine in which a single radiation cavity-field mode acts as the working substance and a single atom mimics an entire non-thermal bath.

Thermalization of the working substance is an important process for any reciprocating quantum heat engine~\cite{quan_quantum_2007}. For instance, if we are able to choose some reservoir parameters to obtain a faster thermalization of the working substance, then, in principle, we could boost the power output of the engine during some finite-time operation~\cite{geva_quantummechanical_1992}.

In this work we investigate how is the quantum thermalization process of a single radiation cavity-field mode in contact with  an effective thermal reservoir made of multi-atoms or multi-level atoms. The theoretical framework of the micromaser model in which the rest of the paper is based on is given in Sec.~\ref{sec2:repeated}. Analytical results for the corresponding thermalization time as well as the numerical simulation of the micromaser are given in Sec.~\ref{sec:3results}. In Sec.~\ref{sec4:applications} we discuss possible implications of our results in the performance of quantum heat engines. We give our conclusions in Sec.~\ref{sec5:conclusions}.

\section{Repeated interaction scheme:\\ Micromaser model}\label{sec2:repeated}

As previously stated, try to define an unambiguous notion of heat and work in the quantum regime is one of the open problems in the emerging field of quantum thermodynamics~\cite{KoslDyna} from which several novel approaches have emerge~\cite{Janet2016}. Among them the general framework of repeated (random or regular) interactions~\cite{strasberg_quantum_2017} has been one in which the first proposals of quantum heat engines appeared~\cite{scully_extracting_2003}. In the field of cavity-QED examples of such a general framework can be found. Cavity-QED has mature methods of quantum control and the micromaser constitutes a good example of a repeated interactions scheme.

\subsection{Injection of two-level atoms}

\begin{figure}
\includegraphics[width=3.2in]{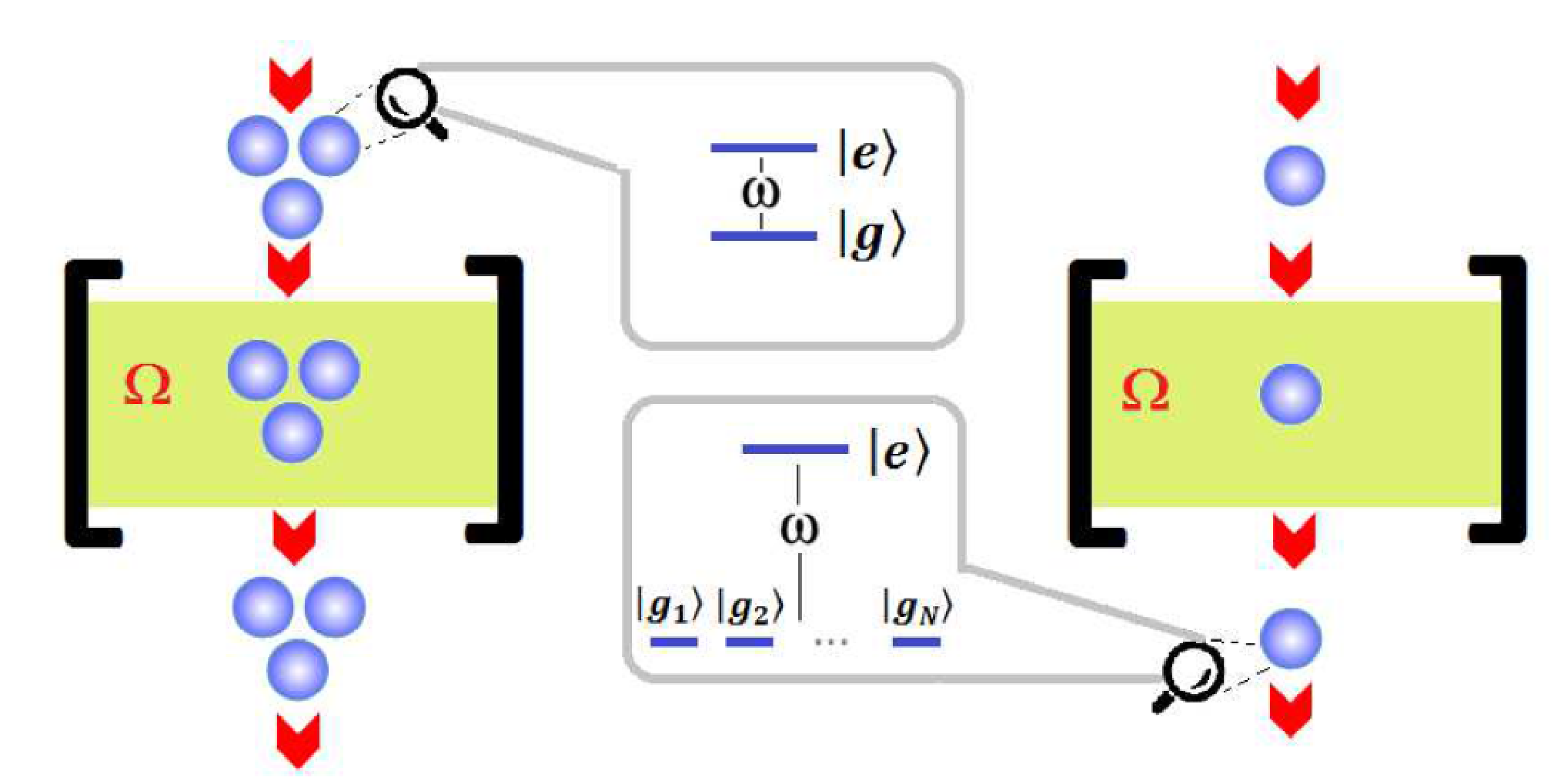}
\caption{(Color online) Schematic setup of two micromaser configurations in which a single radiation cavity-field mode (of frequency $\Omega$) thermalizes due to the fact of repeated random interactions with an atomic beam that traverses  the cavity. The beam is made of (left) clusters of $N$-two-level non-interacting atoms of frequency $\omega$ or (right) a single atom with an excited state $|e\rangle$ and $j$ degenerate ground levels $|g_j\rangle$. Each setup gives a different scaling law for the thermalization time of the cavity-field. Hence, the dynamics of cavity-field can be tuned just by changing the size of the atomic clusters or the number of degenerate ground states.}
\label{fig:Fig1} 
\end{figure}

In a standard micromaser configuration one has a single radiation field mode inside a high reflective electromagnetic cavity and a beam of  atoms crossing it, see Fig.~\ref{fig:Fig1}. The atomic beam consists of uncorrelated velocity-selected-atoms sent from an oven through the cavity with an injection rate $r$. If each atom is approximated by a two-level system its interaction with the cavity-field mode can be described (during a time $\tau$) by the Jaynes-Cummings Hamiltonian \cite{scully_quantum_1997}

\begin{equation}\label{eq:JC}
\hat H=\Omega \hat{a}^{\dagger}\hat{a}+{\omega}\hat\sigma_z/2+g(\hat{\sigma}_{+}\hat{a}+\hat{\sigma}_{-}\hat{a}^{\dagger}),
\end{equation}   
where $\omega$ is the atomic energy level spacing, $\Omega$ the cavity-field frequency and $g$ the atom-field coupling strength (we have set $\hbar=1$). $\hat\sigma_{z}$ and  $\hat\sigma_{\pm}$ are the Pauli-z and Pauli-rising (lowering) operators. $\hat{a}$ and $\hat{a}^{\dagger}$ are, respectively, the annihilation and creation bosonic field operators, they satisfy the usual commutation relation $[\hat{a},\hat{a}^{\dagger}]=1$. 

In this model the cavity decay rate $\kappa$ is negligible because the photon life time $\kappa^{-1}$ is much larger than the time $\tau$ in which the atom-field interaction takes place; a similar argument would apply for neglecting the spontaneous atomic emission. Moreover, the atom-field interaction time is such that $\tau<r^{-1}$ guarantees that only a single atom (or atomic cluster) is inside the cavity each time. 

According to the standard micromaser theory the dynamics of the cavity-field state $\rho_f$ is described, after tracing out the atomic degrees of freedom, by the following coarse-grained master equation~\cite{filipowicz_theory_1986,liao_single-particle_2010}
\begin{equation}\label{eq:Cg}
\dot{\rho}_f= \frac{1}{\tau}\sum_{n=1}^{\infty}\frac{1}{n!}(-i\tau)^n\mbox{Tr}_{a}\left[ \hat{H}_I,
[\hat{H}_I\thinspace ...\thinspace,[ \hat{H}_I,\rho(t)]...\thinspace\right]_n,
\end{equation}
where the dot means the time derivative. $\hat{H}_I$ is the Hamiltonian in the interaction picture and $\mbox{Tr}_a$ is the partial trace operation over the atomic degrees of freedom.
$\rho(t)$ is the density matrix of the overall system so that $\rho=\rho_{a}\otimes\rho_f$ and $\rho_{a}$ is the atomic density matrix. This means that before each interaction the state of the cavity-field has no correlations with the incoming two-level atom.

If the injected two-level atoms are in resonance with the cavity-field the Hamiltonian in the interaction picture is just $\hat H_I=g(\hat\sigma_{+}\hat{a}+\hat\sigma_{-}\hat{a}^{\dagger})$. Usually, it is reasonable to assume that the atoms coming from the oven are in an incoherent superposition between the excited $|e\rangle$ and ground state $|g\rangle$, with probabilities $p_e$ and $p_g$ respectively. In such a case the expression for the two-level atom density operator, in the energy basis, is
\begin{equation}\label{eq:Init_qubit}
\rho_{a}=\left(\begin{array}{cc} p_e & 0\\ 0 & p_g\end{array}\right),
\end{equation}
with $p_e+p_g=1$. If $p_g>p_e$ we can associate a well defined {\em effective} temperature for each two-level atom given by
\begin{equation}\label{eq:Tempat}
T_a=\frac{\omega}{k_B}{\rm ln}\left(\frac{p_g}{p_e}\right)^{-1},
\end{equation} 
where $k_B$ is the Boltzmann constant and the detailed balance condition reads $p_g=\exp[\omega/(k_BT_a)]p_e$. Due to the fact we are not considering any coherences in the state (\ref{eq:Init_qubit}) then~(\ref{eq:Tempat}) coincides with the expressions of the {\em virtual} temperature and the {\em apparent} temperature defined in~\cite{virtual_temperature} and~\cite{apparent_Petruccione} respectively.

With these assumptions the computation of~(\ref{eq:Cg}) up to second order in $n$ yields~\cite{dag_multiatom_2016}:
\begin{align}\label{eq:Lnb}
\dot{\rho}_f=-&\mathcal{R}_a\left(\hat{a}\hat{a}^{\dagger}\rho_f+\rho_f\hat{a}\hat{a}^\dagger-2\hat{a}^{\dagger}\rho_f\hat{a}\right)/2\nonumber\\
&-\mathcal{R}_b\left(\hat{a}^{\dagger}\hat{a}\rho_f+\rho_f\hat{a}^\dagger\hat{a}-2\hat{a}\rho_f\hat{a}^{\dagger}\right)/2,
\end{align}
where $\mathcal{R}_a=\langle\hat\sigma_+\hat\sigma_-\rangle r(g\tau)^2=p_er(g\tau)^2$ and $\mathcal{R}_b=\langle\hat\sigma_-\hat\sigma_+\rangle r(g\tau)^2=p_gr(g\tau)^2$ are known as the excitation and de-excitation rates respectively. Equation~(\ref{eq:Lnb}) is analog to the Markovian master equation commonly used in the theory of open quantum systems to describe the evolution of a quantum harmonic oscillator coupled to a thermal bath, the latter is made of an infinite number of harmonic oscillators. Therefore, the repeated random interaction between the cavity-field mode and each injected two-level atom in a mixed state mimics a thermal bath for the former~\cite{liao_single-particle_2010,strasberg_quantum_2017,apparent_Petruccione}.


Using~(\ref{eq:Lnb}) and the commutation relation of the field operators we can obtain the equation of motion for the  number operator $\hat{n}\equiv\hat a^\dagger\hat a$ given by
\begin{equation}
\dot{\bar{n}}=-(\mathcal{R}_b-\mathcal{R}_a)\bar{n}+\mathcal{R}_a,
\end{equation}
where $\dot{\bar{n}}$ is the rate of change of the average photon number $\bar{n}={\rm Tr}[\rho_f\hat{n}]$. Its time-dependent solution is:
\begin{align}\label{eq:Sol}
\bar{n}(t)=n(0)e^{-(\mathcal{R}_b-\mathcal{R}_a)t}+\frac{\mathcal{R}_a}{\mathcal{R}_b-\mathcal{R}_a}\big[1-e^{-(\mathcal{R}_b-\mathcal{R}_a)t}\big].
\end{align}
The steady state value ${\mathcal{R}_a}/({\mathcal{R}_b-\mathcal{R}_a})$ can be reduced to 
\begin{align}
\frac{\mathcal{R}_a}{\mathcal{R}_b-\mathcal{R}_a}=&\frac{p_e}{p_g-p_e}=\frac{1}{\frac{p_g}{p_e}-1}\nonumber\\
=&({\text{exp}[\omega/(k_BT_a)]-1})^{-1}\equiv\bar{n}_{th},
\end{align}
where the last term is the average thermal photon number. $\bar{n}(t)$ decays from its initial condition $n(0)$ to the stationary value $\bar{n}_{th}$ following the decay rate 
\begin{equation}\label{eq:Gamma}
\Gamma\equiv \mathcal{R}_b-\mathcal{R}_a=r(g\tau)^2(p_g-p_e).
\end{equation}

Equation~(\ref{eq:Sol}) dictates the dynamics of the mean photon number of the cavity-field through a steady state. Within the weak coupling regime an instantaneous effective temperature can be attributed to the cavity-field in terms of the thermal photon number, this is
\begin{equation}\label{eq:Temp}
T_f=({\Omega}/k_B) {\rm ln}\big(1+\bar{n}(t)^{-1}\big)^{-1}.	
\end{equation} 
To maintain the weak coupling regime $g\tau$ should be a small number~\cite{liao_single-particle_2010}. As the system evolves $T_f$ approaches to its stationary value, when this is close to the temperature of the reservoir, defined by Eq.~(\ref{eq:Tempat}), we consider that the cavity-field has thermalized. Hence the temperature of cavity-field in the steady state is always the temperature of the atomic reservoir. This situation changes drastically when quantum coherences~\cite{scully_extracting_2003,PhysRevA.96.032117} or correlations~\cite{dillenschneider_energetics_2009,gerzontemp} in the atomic beam are included, however, in this work we will not address such cases.

We can rewrite $\bar{n}(t)$ as
\begin{equation}
\bar{n}(t)=\bar{n}_0e^{-t/t_{th}}+\bar{n}_{th}(1-e^{-t/t_{th}}),
\end{equation}
where $\bar{n}_0=\bar{n}(0)$ is the initial photon number and $t_{th}$ is the so called thermalization time defined as~\cite{manatuly2018collectively}
\begin{equation}\label{eq:th_Time}
t_{th}\equiv\frac{1}{\Gamma}.
\end{equation}   
With this expression, it becomes possible to evaluate the thermalization time of the cavity-field in terms of some of the atomic beam parameters like population inversion in the case of two-level atoms. In the next subsections, where the beam is made of cluster of atoms or multi-level atoms, we will see that the thermalization time will depend on the cluster's size and the number of lower energy levels respectively.

\subsection{Injection of clusters of $N$ two-level atoms}

Now we consider that clusters of two-level atoms are being injected into the electromagnetic cavity as depicted in the left side of Fig.~\ref{fig:Fig1}. The interaction of $N$ identical, non-interacting, two-level atoms with the cavity-field mode can be described by the Tavis-Cummings Hamiltonian \cite{tavis_exact_1968,tavis_approximate_1969}:

\begin{equation}\label{eq:Tavis}
\hat H=\Omega\hat{a}^{\dagger}\hat{a}+\omega \hat S_z+g\big(\hat{a}^{\dagger}\hat S^{-}+\hat{a}\hat S^{+}\big)
\end{equation} 
where $\hat S_z=\sum_{i=1}^N \hat\sigma_z^i/2$ and $\hat S^{\pm}=\sum_{i=1}^N \hat\sigma_{\pm}^i$. Since the clusters involved are in a thermal non-correlated state, they will also mimic a thermal reservoir for the cavity-field with an effective atomic temperature $T_a$ given by (\ref{eq:Tempat}). The initial state of each atomic cluster $\rho_{a}$ is written by the tensor product of each individual atomic density matrices as

\begin{equation}\label{eq:Cluster}
\rho_{a}=\bigotimes_{j=1}^N
\begin{pmatrix} p_e&0\\0&p_g
\end{pmatrix}.
\end{equation}
Notice that the matrix representation of each interaction Hamiltonian $\hat H_I=g(\hat{a}^{\dagger}\hat S^{-}+\hat{a}\hat S^{+})$ is obtained by straightforward calculations 

\begin{equation}
\hat H_I=g\left(\begin{array}{cc} D&X\\ 		
							   X^{\dagger}&D \end{array}\right)_{2^N\times 2^N}	,			
\end{equation}
where 

\begin{equation}
D=\left(\begin{array}{ccccc}  &\hat{a}&\cdots&\hat{a}&\\
							  \hat{a}^{\dagger}& & & &\hat{a}\\
							  \vdots& & & &\vdots\\	
							  \hat{a}^{\dagger}& & & &\hat{a}\\
							   &\hat{a}^{\dagger}&\cdots&\hat{a}^{\dagger}&
							  	
							  \end{array}\right)_{2^{N-2}\times 2^{N-2}}
\end{equation}
is the block diagonal part.

\begin{equation}
X=\left(\begin{array}{ccc}  \hat{a}& & \\
					        & \ddots & \\
					        & & \hat{a}
					        \end{array}\right)_{2^{N-2}\times 2^{N-2}}
\end{equation}
 is the inverse diagonal part of $\hat H_I$.
%
%
Inserting these expressions of $\rho_a$ and $\hat H_I$ in~(\ref{eq:Cg}), and following all the steps described in previous subsection, the master equation of the cavity-field mode can be obtained:
\begin{align}\label{eq:Master1}
\dot{\rho}_f=r\phi^2N p_g \mathcal{L}[\hat{a}]+r\phi^2N p_e\mathcal{L}[\hat{a}^{\dagger}],
\end{align}
where $\phi=g\tau$, $N$ is the number of two-level atoms of the injected multi-atom cluster and $\mathcal{L}[x]=\frac{1}{2}(2x\rho x^{\dagger}-x^{\dagger}x\rho-\rho x^{\dagger}x)$ is the well known Liouvillian superoperator in the Lindblad form. To obtain~(\ref{eq:Master1}) we also made use of the fact that $\langle\hat S_-\hat S_+ \rangle =N p_g$  and $\langle\hat S_+\hat S_- \rangle =Np_e$; these expectation values were calculated with respect of the atomic cluster state~(\ref{eq:Cluster}). 
The dependency of $N$ in the above equation is evident and it will strongly affect the dynamics of the cavity-field as we will discus in the next section. If $N=1$ ~(\ref{eq:Master1}) reduces to (\ref{eq:Lnb}) as expected.

\subsection{Injection of multi-level atoms}

On the other hand, the interaction of the cavity-field with a single atom having one excited state $|e\rangle$ and $i$ number of lower energy levels $|g_i\rangle$ (see right side of Fig.~\ref{fig:Fig1}) can be described by \cite{liao_single-particle_2010}:
\begin{eqnarray}\label{eq:ham:mult:levl}
\hat H=\Omega \hat{a}^{\dagger}\hat{a}+\omega_e |e\rangle\langle e|+{\small \sum}_{i=1}^N\omega_{b_i}|g_i\rangle\langle{g_i}|
+g\big(\hat{R}_{+}\hat{a}+\hat{R}_{-}\hat{a}^{\dagger}\big)\nonumber\\
\end{eqnarray}
where $\hat{R}_{+}=\frac{1}{\sqrt{N}}\sum_{i}^N |{e}\rangle\langle{g_i}|$ and $\hat{R}_{-}={R}_{+}^{\dagger}$. 
The coupling of all the atomic energy levels with field is equal, besides all the lower levels are taken as degenerate ground states with $\omega_{b_i}=0$. Also, for simplicity, we assume that all $|g_i\rangle$ are equally populated, thus conservation of the total probability of a multi-level atom is $p_e+N p_{g^{'}}=1$, where $p_{g^{'}}$ is the population of one of the degenerate ground states. 

In the energy basis the state of each of these multi-level atoms can be defined by the following diagonal matrix 
\begin{equation}
\rho_{a}=\left(\begin{array}{cccc}  p_e & & &\\
									&p_{g^{'}}& & \\
									 & &\ddots & \\
									 & & & p_{g^{'}}  			 
\end{array}\right)_{N+1\times N+1}.
\end{equation} 
In this case, the associated effective temperature of the atomic reservoir is $T_a=(\omega/k_B)\big( {\rm ln}(Np_{g^{'}}/p_e)\big)^{-1}$. 
\begin{figure*}[!t]
\includegraphics[width=6.74in]{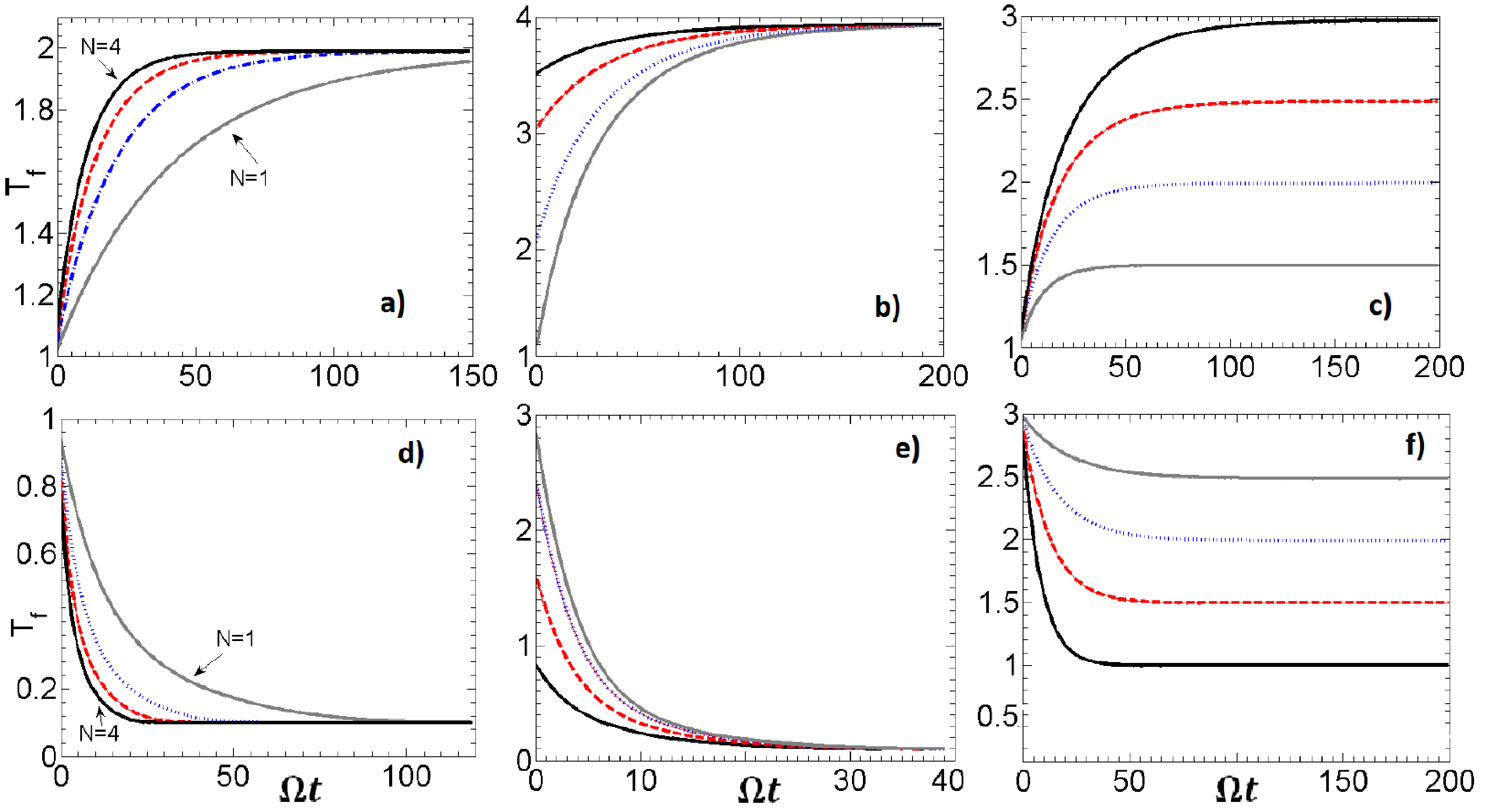}
\caption{(Color online) Numerical results for the simulation of the micromaser configuration showed in the left side of Fig.~\ref{fig:Fig1}. Panels show, using the effective temperature $T_f$, the thermalization process of a single cavity-field mode due to the repeated interaction with clusters of $N$ two-level atoms in a thermal state of temperature $T_a$. Upper (lower) row shows a {\em heating} ({\em cooling}) process in which the initial $T_f$ is smaller (larger) than $T_a$. a) and d) show a reduction in the thermalization time when the number of atoms $N$ increases from $1$ to $4$; initial values are $T_f=1$ and, $T_a=2$ and $T_a=0.1$ for a) and d) respectively. b) [also $e)$] shows $T_f$ for four different initial values and $T_a=4$ (0.1). Finally, c) and f) show $T_f$ for four different values of $T_a$. The coupling strength is $g=0.08$ for a) and d), and $g=0.1$, $N=2$ for the rest of the panels. Atomic and field decay rates are $\gamma=10^{-9}$ and $\kappa=0.5\times10^{-10}$. $T_f$ ($T_a$) is in units of $\Omega/k_{B}$ ($\omega/k_B$). The $x$ axis is the scaled time and it represents the number of interactions between the cavity-field and the atomic beam.} 
\label{fig:Fig2}
\end{figure*}
The Hamiltonian~(\ref{eq:ham:mult:levl}) in the interaction picture has a matrix representation
\begin{equation}
\hat H_I=\frac{g}{\sqrt{N}}\left(\begin{array}{cccc}   &\hat{a}&\cdots&\hat{a}\\
														\hat{a}^{\dagger}& & & \\
									 					\vdots& & \\
														 \hat{a}^{\dagger}& & &   			 
\end{array}\right)_{N+1\times N+1},
\end{equation} 
where $N$-ground energy levels have been considered. Following the same steps as in the previous subsections the master equation for the cavity-field can be derived:
\begin{eqnarray}\label{eq:Master2}
\dot{\rho}_f=r\phi^2 p_{g^{'}} \mathcal{L}[\hat{a}]+r\phi^2 p_e \mathcal{L}[\hat{a}^{\dagger}].
\end{eqnarray}
In this case we use the fact that $\langle \hat R_- \hat R_+\rangle=p_g'$ and $\langle \hat R_+ \hat R_-\rangle=p_e$.

\section{Results}\label{sec:3results}

\subsection{Thermalization time}

Master equations~(\ref{eq:Master1}) and (\ref{eq:Master2}) have the same structure of~(\ref{eq:Lnb}), hence we can use the result of~(\ref{eq:th_Time}) where $t_{th}\equiv\Gamma^{-1}$, to compute the thermalization time for the cases in which the atomic beam is made of clusters of two-level atoms as well as multi-level atoms. For the atomic cluster reservoir the associated thermalization time is
\begin{equation}\label{eq:T1}
\mathcal{T}_{\rm cl}\equiv\frac{1}{r\phi^2N(p_g-p_e)}.
\end{equation}
Taking the low temperature condition $p_e\ll p_g$ for the atomic cluster beam, (\ref{eq:T1}) becomes $\mathcal{T}_{\rm cl}\approx 1/(r\phi^2N)$. Therefore, with a large number of two-level atoms in the cluster one can reduce the thermalization time. A similar expression to (\ref{eq:T1}) was obtained in~\cite{manatuly2018collectively} when the cavity-field is replaced just by a single qubit, the differences comes from the intrinsic symmetry characterizing each system. In contrast, for the multi-level atomic reservoir, the thermalization time is 
\begin{equation}\label{eq:T2}
\mathcal{T}_{\rm ml}\equiv\frac{1}{r\phi^2(p_{g^{'}}-p_e)}.
\end{equation}
Since $p_{g^{'}}=(1-p_e)/N$ we obtain $\mathcal{T}_{\rm ml}\approx N/(r\phi^2)$ in the low temperature limit. The multi-level atomic reservoir increases $\mathcal{T}_{\rm ml}$ following a scaling linear law with respect to $N$. Thus, these results show that the multi-atom or multi-level atomic reservoirs might be used to reduce or delay, respectively, the corresponding thermalization time of the cavity-field mode on demand. 

\subsection{Numerical simulation}
 
Up to this point, using an approximate coarse-grained master equation, the cavity-field dynamics was obtained. Also, in addition to the weak-coupling regime assumption, all internal decoherent channels of each subsystem, cavity-field and atoms, were ignored. Now, we will perform a numerical simulation of the injection of  each type of atomic reservoir through the cavity along with their interaction with the radiation field. In addition, cavity and atomic decay rates will be considered. 

In a standard micromaser configuration the process of atomic injection can be characterized by an injection rate defi\-ned by $r=1/(\tau+\tau_0)$, where $\tau$ is the duration of the matter-field interaction and $\tau_0$ is the time elapsed when there is no matter present in the cavity, i.e., the time before the next atomic element enters into the cavity. The initial state of the total system is $\rho=\rho_a\otimes\rho_f$, this state is updated right after each matter-field interaction takes place. 

For the multi-atom reservoir case we will describe the evolution of the total system (during each interaction time) by the following Markovian master equation:
\begin{equation}\label{eq1:mas:num}
\dot{\rho}=-i[\hat H,\rho]+\gamma{\small\sum}_i\mathcal{L}[\hat\sigma_{-}^i]+\kappa\mathcal{L}[\hat{a}],
\end{equation}
where $\hat H$ is total system Hamiltonian (\ref{eq:Tavis}), $\gamma$ and $\kappa$ are the atomic and cavity decay rates. In contrast with (\ref{eq:Cg}) and to be more realistic the above equation takes into account possible energy losses. As we are interested in the evolution of the average photon number $\bar{n}$, the instantaneous effective cavity-field temperature (\ref{eq:Temp}) and the corresponding thermalization time, we need to perform in (\ref{eq1:mas:num}) the partial trace out of the atomic degrees of freedom. Doing this the cavity-field state will evolve iteratively through a steady state dictated by the multi-atom reservoir (\ref{eq:Cluster}).

We set the initial state of the cavity-field mode to be the thermal state $\rho_f(0)=\frac{1}{Z}e^{-\beta \hat H_f}$, with $\hat H_f=\Omega\hat{a}^{\dagger}\hat{a}$, $\beta=(k_B T_f)^{-1}$ and $Z={\rm Tr}[e^{-\beta \hat H_f}]$ the partition function. During the calculations $\rho$ will be represented by a finite square matrix due to the fact that we have to truncate the corresponding Hilbert space of the cavity-field state for any numerical implementation. For simplicity all two-level atoms in the cluster are assumed to be resonant with the cavity-field with the identical frequencies $\omega_1=...=\omega_N=\Omega$. 
For practical purposes and without loss of generality we assume, in all the numerical calculations, a regular atomic injection through the cavity and $\tau_0=0$~\cite{apparent_Petruccione}. In the following we will see that previous analytical results given by the coarse-grained master equation (\ref{eq:Master1}) are in good agreement with the numerical calculations.

Figure~\ref{fig:Fig2} shows the effective temperature $T_f$ of the cavity-field as a function of the number of interactions with the multi-atom reservoir. As expected, after a finite number of interactions $T_f$ reaches the effective temperature $T_a$ of the atomic-reservoir. However, if the size of the atomic clusters increase, a significant reduction in the number of interactions, needed to converge the thermal value, is observed [see Fig.~\ref{fig:Fig2} a) and d)]; this is in agreement with the result predicted by (\ref{eq:T1}). Fi\-gu\-re~\ref{fig:Fig2} b) [Fig.~\ref{fig:Fig2} e)] shows the evolution of $T_f$ for four different initial values smaller (larger) than the effective temperature of the atomic cluster, i.e., it represents a ``heating" (``cooling") process. Figs.~\ref{fig:Fig2} c) and f) show the behavior of $T_f$ when four different effective temperatures of the atomic cluster are considered during a heating and cooling process respectively. In this examples the low temperature condition $p_e\ll p_g$ was also considered. 

If the multi-level atoms are injected into the cavity the evolution of the total system will be described by 
\begin{equation}\label{eq2:mas:num}
\dot{\rho}=-i[\hat H,\rho]+\gamma{\small \sum}_m^{N+1}\mathcal{L}[\hat{R}_{-}^m]+\kappa\mathcal{L}[\hat{a}],
\end{equation}      
where $\hat H$ is given in (\ref{eq:ham:mult:levl}) and $\hat{R}_{-}^m=|{g}\rangle\langle{\alpha_m}|$ with $\alpha_m=e,g_1,\ldots g_N$. The respective numerical calculations were performed following the same arguments as previous case.

\begin{figure*}[!t]
\includegraphics[width=6.74in]{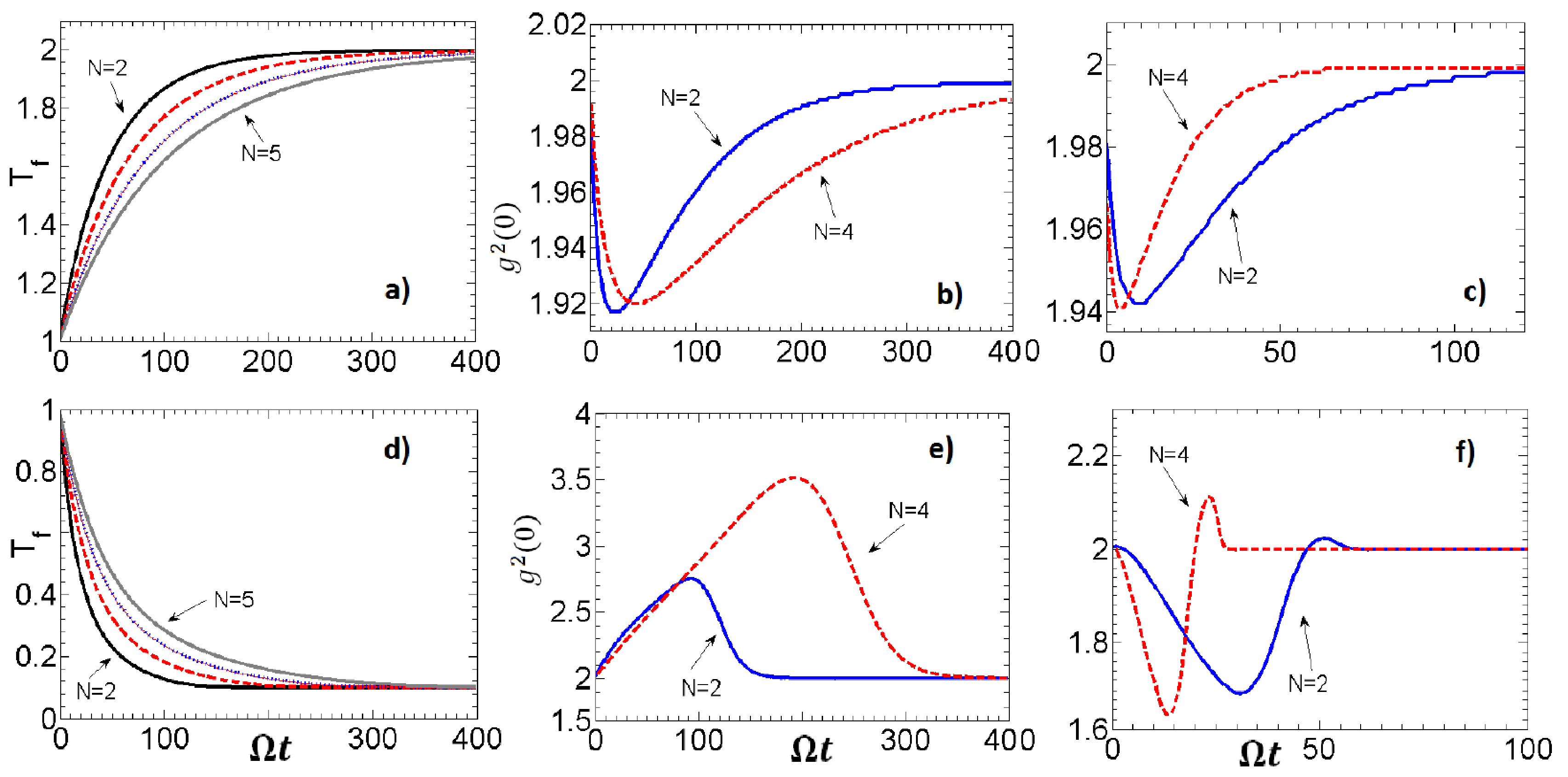}
\caption{(Color online) Time evolution of the cavity-field temperature $T_f$ and the second order correlation function $g^2(0)$ when multi-level atoms, with different numbers of ground energy levels, are injected into the cavity, see right side of Fig.~\ref{fig:Fig1}. Subplots a), b) and c) depict a ``heating" process of the cavity-field while d), e) and f) a ``cooling" one. In units of $\Omega/k_B$ the initial field temperature is $T_f=1$ for both a) and d). In units of $\omega/k_B$ the temperature $T_a$ of each injected multi-level atom is $2$ and $0.1$ for a) and d) respectively. The atom-field coupling $g=0.1$, spontaneous emission coefficient $\gamma=10^{-9}$ and the cavity decay rate $\kappa=0.5\times10^{-10}$ are the same for all subplots. The $x$ axis is the scaled time and it represents the number of interactions between the cavity-field and the atomic reservoir. For more details see the main text.} 
\label{fig:Fig3}
\end{figure*}

Figures~\ref{fig:Fig3} a) and d) show the numerical evolution of the field temperature $T_f$. There the thermalization time of the cavity-field is enlarged if the number of the degenerate ground states of the atomic beam increase; they represent a ``heating" and ``cooling" process respectively.

On the other hand, to elucidate what is the nature of the final state of the system, we performed a numerical evaluation of the second order correlation function $g^{(2)}(\tau)$ of the cavity-field~\cite{gerry2005introductory}. $g^{(2)}(\tau)$ is used to characterize the photon statistics of classical and non-classical sources of ligh, at zero time delay it is defined by $g^2(0)=\langle \hat{a}^{\dagger 2}\hat{a}^2\rangle/\langle \hat{a}^{\dagger}\hat{a}\rangle^2$~\cite{scully_quantum_1997}. For a thermal (coherent) state $g^2(0)=2$~($1$) while for a squeezed thermal state it is larger than $2$~\cite{kim_properties_1989}.

Figure~\ref{fig:Fig3} [sets b), c), e) and f)] shows the numerical results of $g^2(0)$ as function of the number of interaction with the atomic beam reservoir that is made of cluster of two-level atoms [b) and e)] as well as multi-level atoms [c) and f)]. We can observe that $g^2(0)$ converges to $2$ in the steady state, thus the cavity-field state evolved in a thermal state. 

\section{Application for quantum heat engines}\label{sec4:applications}

Atomic reservoirs come out by different ways in physical systems. Ultracold atoms can be a convenient example of atomic reservoirs.  For instance, to describe non-equilibrium atomic transport they can be in contact with the edge-sites of a finite one-dimensional lattice of Bose atoms~\cite{kordas_non-equilibrium_2015}. These kind of setups are recently used in atomtronics, i.e., cold-atoms simulate the components of electronic devices~\cite{kolovsky_microscopic_2017}. On the other hand, atomic reservoirs can be used as effective thermal baths for reciprocating quantum heat engines exploiting the light-matter interaction. A photonic-Carnot engine~\cite{turkpence_quantum_2016,turkpence_photonic_2017,scully_extracting_2003} operating with a single radiation cavity-field mode as the working substance is a good example. There, the radiation pressure plays the role of a piston driving one of the high-reflective (movable) cavity mirrors while the other (fixed) is in contact with an ordinary thermal bath.  

In general, quantum heat engines operate using classical thermodynamic cycles while the working substance is replaced by a quantum system~\cite{quan_quantum_2007}. For instance, a quantum-Otto cycle consists of four strokes: isentropic compression, hot isochoric stage, isentropic expansion, and cold isochoric stage.
During the isochoric stages the quantum system is in contact with a reservoir where heat exchange occurs and no work is done during the process. These two strokes are the quantum thermalization parts of the cycle and end up (normally) with a thermal quantum state of the working substance. The isentropic strokes are the quantum adiabatic processes in which a unitary evolution takes place and the system Hamiltonian has time-dependent parameters, therefore work is done during these stages~\cite{KoslDyna}. In the adiabatic strokes there is no heat transfer between the system and the environment because the working substance is disconnected from the reservoir. 

The efficiency and power of a quantum heat engine can be expressed, respectively, in terms of their conventional definitions, as the ratio of the work output and work input and as the work output per unit time respectively. Ideally, the quantum adiabatic strokes are infinitely slow and hence the power vanish. The power output should be a quantifier in a finite-time thermodynamic quantum cycle~\cite{geva_quantummechanical_1992,feldmann_performance_2000}; in a finite-time thermodynamic quantum cycle adiabatic strokes also operate at a finite-time. It is a straightforward inference to expect high power output using a shorter cycle time. However, in finite-time thermodynamics increasing the efficiency generally yields a power decrease and vice versa~\cite{feldmann_performance_2000}. Unless one makes a shortcut to adiabaticy~\cite{PhysRevLett.104.063002,torrontegui2013shortcuts,PhysRevLett.118.100601,ccakmak2018spin}, during a rapid adiabatic drive unwanted entropy production is present and internal friction occurs due to the fact of the non-commuting Hamiltonians at different instants of time~\cite{feldmann_quantum_2006, plastina_irreversible_2014,cakmak_irreversibility_2016}. Since these adiabatic effects limit the performance of quantum thermal machines, shortening the quantum isochoric strokes of a cycle (where quantum thermalization occurs) becomes important in order to enhance their performance. 
Therefore, the results of this work can be used to boost the performance of thermal machines through reducing their corresponding thermalization time. Though isochoric paths of the cycles are (normally) much shorter than the adiabatic paths, any shortening of thermalization time will directly be translated to the improvement of the performance of the cycle without affecting the work output since no work is done on these paths. Also, our results would be beneficial in open quantum systems demanding longer thermalization times. 

\section{Conclusions}\label{sec5:conclusions}

Using a repeated interaction scheme, the micromaser model, we investigated the quantum thermalization dynamics of a bosonic cavity-field mode when interacts with two types of atomic reservoirs. These consisted of cluster of $N$ two-level atoms as well as multi-level atoms having $N$ degenerate ground energy levels. Depending on the type of atomic reservoir analytic expressions for the corresponding thermalization time of the cavity-field [see (\ref{eq:T1}) and (\ref{eq:T2})] were obtained. For the multi-atom reservoir case we found that the thermalization time is inversely proportional with the number of two level atoms~(\ref{eq:T1}). In contrast, it is proportional to the number of degenerate energy levels in the multi-level reservoir case~(\ref{eq:T2}). Performing a numerical simulation for the injection of each atomic reservoir into the cavity, along with the numerical integration of the microscopic master equations [(\ref{eq1:mas:num}) and (\ref{eq2:mas:num}], the analytical results were verified. We also discussed how tailoring the thermalization time of a quantum system might be always useful for applications in open quantum systems. For instance, in quantum thermodynamics, if such a quantum system is considered as the working substance of a quantum heat engine, then a shortening in its thermalization time could boost the performance and power output of the engine.

\acknowledgments

D.T. acknowledges support from \.{I}stanbul Technical University.  D.T. specially thanks to Ferdi Altintas for guiding discussions. R. R.-A. acknowledges support from University Research Agreement between Ko\c{c} University and Lockheed Martin Chief Scientist’s
Office.



%

\end{document}